 \documentclass[final,3p,times]{elsarticle}
\usepackage{amsmath,hyperref}
\usepackage{lineno}

\usepackage{epstopdf}
\journal{XXX}

\usepackage{lineno,hyperref}
\usepackage{graphicx}
\usepackage{dcolumn}
\usepackage{bm}
\usepackage{epsfig}
\usepackage{booktabs}
\usepackage{subfigure}
\usepackage{graphics}
\usepackage{amssymb}
\usepackage{amsmath}
\usepackage{array}
\usepackage{color}
\usepackage{booktabs}
\usepackage{multirow}
\usepackage{caption}
\usepackage{chngpage}
\usepackage{subfigure}
\usepackage{mathrsfs,gensymb,float}

\biboptions{numbers,sort&compress}
\modulolinenumbers[1]

\begin{document}

\captionsetup[figure]{labelfont={bf},name={Fig.},labelsep=period}        

\begin{frontmatter}
	
	\title{Droplet impact dynamics on Janus-textured heated substrates}
	
	\author[1,2]{Jiangxu Huang}
	\author[1,2]{Lei Wang\corref{mycorrespondingauthor}}
	\cortext[mycorrespondingauthor]{Corresponding author}
	\ead{leiwang@cug.edu.cn}

	\address[1]{School of Mathematics and Physics, China University of Geosciences, Wuhan 430074, China}
	\address[2]{Center for Mathematical Sciences, China University of Geosciences, Wuhan 430074, China}
	

	
\begin{abstract}

In this paper, the droplet impact dynamic behavior of droplet on Janus-textured heated substrates is numerically investigate by using the thermal Lattice Boltzmann method. The effect of several factors like the wettability, the Jakob number and the Weber number on the droplet impact dynamic behavior on Janus-textured heated surface are studied in detail. The simulation results show three types of boiling states, i.e., contact boiling, transition boiling and film boiling. The droplets in contact boiling state directionally migrate toward the denser region, while the droplets move toward the sparser region for film boiling state at low Weber number, which agrees with the direction of droplet motion in experiments [Zhang et al., Adv. Funct. Mater., 2019, 29, 1904535]. The self-propelled droplets on the Janus-textured heated substrates is owning to unbalanced Young's fore, vapor pressure difference and thermophoretic force. Moreover, The Janus-textured surface with low wettability has superior nucleation behavior resulting in their typically easy transition from efficient transition boiling state to inefficient film boiling state. In addition, the boiling states of the droplet change from the contact boiling state to the transition boiling state and then to the film boiling state as the Jakob number increases.

\end{abstract}
	
	
\end{frontmatter}

\section{Introduction}

The impaction of droplets on the heated surface widely exists in many areas of applied science and engineering technologies, such as spray cooling \cite{VisariaITCPT2009}, desalination\cite{ShenDWT2013} and metallurgical treatment\cite{FardIJHFF2001}. When the liquid droplets impact a surface with a temperature around the liquid boiling point, the droplets quickly boil and evaporate. However, when the temperature is far above the liquid boiling point, the droplet may be levitated above the vapor layer caused by the violent boiling beneath the droplet \cite{QuereANFM2013}. Compared with the impinging droplets on the non-heated surface, the mass-heat transfer between the droplet, the heated surface and the ambient environment influences the dynamic behaviours of droplet impact. Due to its application value in many practical engineering, the investigation of the impinging droplet on the heated surface has attracted considerable attention from the scientific community.

Since the pioneering experiments of droplet impact on a hot wall conducted by Wachters et al. \cite{WachtersCES1966}, a considerable number of experimental and numerical investigations of droplet impingement on heated substrates have been performed \cite{BernardinIJHMT1996,KarlPOF2000,GeIJHM,TranPRL2016,KompinskyCES2013,LeePRF2020,ChanteloJFM2021,XuATE2022}. In 1997, Bernardin et al. \cite{BernardinIJHMT1996} used high-speed photography to capture the impact behavior of water droplets on a hot aluminum surface and show that the Weber number has a strong influence on the dynamic behavior of droplets. Later, Karl and Frohn \cite{KarlPOF2000} experimentally studied the effect of impact angle on droplet breakup and found the minimum impact angle of droplet breakup. Then, Ge and Fan \cite{GeIJHM} used the level set method to simulate the sub-cooled droplet impact on the heated surface and found that the Leidenfrost degree decreases as the increasing sub-cooled degree increase. After that, Tran et al. \cite{TranPRL2016} experimental investigated the impinging droplet on the superheated surfaces and showed that due to the evaporation steam ejecting radially outward and taking away the liquid, the maximum spreading factor is proportional to $ We^{2/5} $, which is much larger than for the impact on non-heated surfaces. Kompinsky et al. \cite{KompinskyCES2013} presented an experimental study of binary fuel droplet impacts on a heated surface. They identified a critical surface temperature needed for a droplet breakup, which is independent of impact velocity within the examined range. Recently, Lee et al. \cite{LeePRF2020} reported the downward ejection of Leidenfrost droplets due to the convergence of capillary waves along the liquid cavity interface in the process of impinging on a high-temperature wall, and they proposed a jetting criterion based on the viscous damping of capillary waves. Chantelot et al. \cite{ChanteloJFM2021} experimentally studied the impact of volatile drops on superheated substrates and revealed the influences of the gas drainage and evaporation on the dynamic Leidenfrost transition. More recently, Xu et al. \cite{XuATE2022} used the lattice Boltzmann (LB) method to study the droplet impingement on a heated surface, and the numerical results show three different rebound modes due to the nucleation, growth and rupture of the bubble.

The above research mainly focuses on the impact dynamics on heated smooth surface, the asymmetrical structural surface have attracted extensive attention due to the ability to control the directional movement of droplets. In order to understand the impact and rebound dynamics of droplets on the asymmetrical structural heated surface, several investigations have been conducted to understand the dynamics behavior of droplets impacting the asymmetrical structural heated surface \cite{LinkePRL2006,AgapovACSN2014,LiNP2016,ChenASS2018,ZhaoSCT2018,ZhangAFM2019,LiuAM2020}. In 2006, Linke et al. \cite{LinkePRL2006} firstly reported the self-propelled Leidenfrost droplets driven by the thermal gradient on ratchet surfaces. After that, Agapov et al. \cite{AgapovACSN2014} experimentally investigated the droplet impact on the heated surface with tilted nanopillar arrays and they found that the droplet in the transition boiling regime will directional rebound since asymmetric wettability of the nanostructured surfaces. Then, Li et al. \cite{LiNP2016} experimentally investigated droplet impingement on the Janus-textured heated surface and found that droplets migrate to the region of higher heat transfer coefficient due to two concurrent thermal states breaking the wetting symmetry. Chen et al. \cite{ChenASS2018} explored the self-propulsion of Leidenfrost droplets on hot micro-pillared surfaces with gradient wettability and they found Leidenfrost droplets self-propelled from the smaller superficial area of micropillars towards the larger superficial area of micropillars. Zhao et al. \cite{ZhaoSCT2018} carried out the experiment of droplets impact on heated surface with gradient grooves, the results show that the droplets rebound directionally under the action of the unbalanced Young's force and the vapor pressure difference. Recently Zhang et al. \cite{ZhangAFM2019} constructed a series of Janus-textured substrates to investigating the directional migration of droplet, they found that the Leidenfrost droplet always move toward region with low roughness. More recently, Liu et al. \cite{LiuAM2020} reported the droplet impact on the hot surface with Janus-mushroom structure and it is revealed that the liquid droplets transport directionally due to the directional liquid penetration and vapor flow toward the straight sidewall at high temperatures.

These previous studies demonstrate that the dynamic behavior of droplets impinging on the surface is significantly influenced by the Janus-textured structure. To the best of our knowledge, relatively few numerical investigations have been conducted on an impinging droplet on a surface with a Janus texture. The existing numerical research of impinging droplets was focused on the ratchet surfaces and uniform structural surfaces, e.g., Li et al. \cite{LiSM2016} firstly used the LB method to simulate the self-propelled Leidenfrost droplets on ratchet surfaces. More recently, Xu et al. \cite{XuATE2022} and Wang et al. \cite{WangPOF2022} numerically investigated the droplet impact on the substrate with uniform micro-structured surfaces. Framed in this general background, the dynamics for droplets impacting the Janus-textured heated surface are investigated from the numerical point of view in the present work. The method used in this work is the so-called pseudopotential LB method \cite{ShanPRE1993}, which has been widely used in previous studies of droplet impact on the heated substrates due to its remarkable computational efficiency, clear representation of the underlying microscopic physics, simplicity and versatility \cite{XuATE2022,LiSM2016,WangPOF2022}.

The remainder of the paper is organized as follows. In the following section, the pseudopotential LB model is introduced. Section 3 describes the problem of droplet impact on the heated Janus-textured surface. In Section 4, the pseudopotential LB model is validated with three tests. The numerical results and discussion are presented in Section 5. Finally, we provide a brief conclusion in Section 6.
 
\section{Model description}

\subsection{The pseudopotential multiphase LB method}

The pseudopotential multiphase LB model proposed by Shan and Chen is a particularly popular method in the LB community to deal with multiphase flows \cite{ShanPRE1993}. To achieve thermodynamic consistency and large density ratio, various improved pseudopotential multiphase LB models are proposed in recent years. In the present work, a two-dimensional nine-velocity (D2Q9) multiple-relaxation-time (MRT) improved pseudopotential LB model proposed by Li et al. \cite{LiPRE2012} is adopted,  and the LB equation to describe the density evolution is written as
\begin{equation}
\begin{aligned}
f_\alpha\left(\mathbf{x}+\mathbf{e}_\alpha \delta_t, t+\delta_t\right)= f_\alpha(\mathbf{x}, t)-\left.\bar{\Lambda}_{\alpha \beta}\left(f_\beta-f_\beta^{\mathrm{eq}}\right)\right|_{(\mathbf{x}, t)} 
+\left.\delta_t\left(S_\alpha-0.5 \bar{\Lambda}_{\alpha \beta} S_\beta\right)\right|_{(\mathbf{x}, t)},
\label{eq1}
\end{aligned}
\end{equation}
where $ f_\alpha(\mathbf{x}, t) $ is the density distribution function with velocity at spatial position $\mathbf{x}$ and time $t$, $ f_{\alpha}^{eq} $ is the equilibrium distribution function, $\mathbf{e}_\alpha$ is the discrete velocity along $\alpha$th direction, $\delta$t is the time step, $S_{\alpha}$ is the forcing term in the velocity space, and $\bar{\Lambda}_{\alpha \beta}=\left(\mathbf{M}^{-1} \boldsymbol{\Lambda} \mathbf{M}\right)_{\alpha \beta}$ is collision matrix, in which $\mathbf{M}$ is the transport matrix and $\boldsymbol{\Lambda}$ is a diagonal matrix given by
\begin{equation}
\boldsymbol{\Lambda}=\operatorname{diag}\left(\tau_\rho^{-1}, \tau_e^{-1}, \tau_{\varsigma}^{-1}, \tau_j^{-1}, \tau_q^{-1}, \tau_j^{-1}, \tau_q^{-1}, \tau_v^{-1}, \tau_v^{-1}\right).
\end{equation}
Through the transport matrix $\mathbf{M}$, the right side of Eq. \ref{eq1} can be  implemented in the moment space  
\begin{equation}
\mathbf{m}^*=\mathbf{m}-\boldsymbol{\Lambda}\left(\mathbf{m}-\mathbf{m}^{\mathrm{eq}}\right)+\delta_t\left(\mathbf{
}-\frac{\mathbf{\Lambda}}{2}\right) \mathbf{S},
\label{eq3}
\end{equation}
where $\mathbf{m}=\mathbf{M f}, \mathbf{m}^{\mathrm{eq}}=\mathbf{M} \mathbf{f}^{\mathrm{eq}}$, and $ \mathbf{S} $ is the forcing term in the moment space. After the collision step in the moment space, the  $ \mathbf{m}^* $ can be transformed back to the discrete velocity space. The streaming process in velocity space is then given by
\begin{equation}
f_\alpha\left(\mathbf{x}+\mathbf{e}_\alpha \delta_t, t+\delta_t\right)=f_\alpha^*(\mathbf{x}, t),
\end{equation}
where $\mathbf{f}^*=\mathbf{M}^{-1} \mathbf{m}^*$.  The corresponding equilibrium moments $ \mathbf{m}^{\mathrm{eq}} $ are given by
\begin{equation}
\begin{aligned}
\mathbf{m}^{\mathrm{eq}}= \rho\left(1,-2+3|\mathbf{u}|^2, 1-3|\mathbf{u}|^2, u_x,-u_x, u_y\right.
\left.-u_y, u_x^2-u_y^2, u_x u_y\right)^T.
\end{aligned}
\end{equation}

The forcing term in Eq. \ref{eq3} is taken as follows
\begin{equation}
{\mathbf{S}}=\left[\begin{array}{c}
0 \\
6\left(u_x F_x+u_y F_y\right)+\frac{12 \phi|\mathbf{F}|^2}{\psi^2 \delta_t\left(\tau_e-0.5\right)} \\
-6\left(u_x F_x+u_y F_y\right)-\frac{12 \phi|\mathbf{F}|^2}{\psi^2 \delta_t\left(\tau_{\varsigma}-0.5\right)} \\
F_x \\
-F_x \\
F_y \\
-F_y \\
2\left(u_x F_x-u_y F_y\right) \\
\left(u_x F_y+u_y F_x\right)
\end{array}\right],
\end{equation}
where $\phi$ is a constant that adjusts mechanical stability to achieve thermodynamic consistency. The macroscopic density and velocity are calculated by 
\begin{equation}
\rho=\sum_\alpha f_\alpha, \quad \rho \mathbf{u}=\sum_\alpha \mathbf{e}_\alpha f_\alpha+\frac{\delta_t}{2} \mathbf{F},
\end{equation}
where $ \mathbf{F} $ is the total force term acted on the system, including the liquid-liquid interaction force $ \mathbf{F_m} $, the liquid-solid interaction force $ \mathbf{F}_{\mathrm{ads}} $ and the body force $ \mathbf{F}_{\mathrm{b}} $ \cite{ShanPRE1994,LiPRE2014}.
\begin{equation}
\mathbf{F}_{\mathbf{m}}(\mathbf{x})=-G \psi(\mathbf{x}) \sum_\alpha w \left(\left|\mathbf{e}_\alpha\right|^2\right) \psi\left(\mathbf{x}+\mathbf{e}_\alpha\right) \mathbf{e}_\alpha,
\end{equation}
\begin{equation}
\mathbf{F}_{\mathrm{ads}}(\mathbf{x})=-G_w \psi(\mathbf{x}) \sum_\alpha w \left(\left|\mathbf{e}_\alpha\right|^2\right) \psi(\mathbf{x})s\left(\mathbf{x}+\mathbf{e}_\alpha\right) \mathbf{e}_\alpha,
\label{eq9}
\end{equation}
\begin{equation}
\mathbf{F}_{\mathrm{b}}(\mathbf{x})=\rho(\mathbf{x}) \mathbf{g},
\end{equation}
where G is the interaction strength, $ G_w $ is the adsorption parameter to control the surface wettability, $ s $ is an indicator function, $ \mathbf{g}=(0,-g) $ is the gravitational acceleration, $ w \left(\left|\mathbf{e}_\alpha\right|^2\right) $ are the weight, which are given as $ \omega(1)=1/3$ and $ \omega(2)=1/12$, $ \psi(\mathbf{x}) $ is the interaction potential, which can be obtained by introducing a non-ideal equation of state (EOS) \cite{HeJSP2002,YuanPOF2006}
\begin{equation}
\psi(\mathbf{x})=\sqrt{\frac{2\left(P_{\mathrm{EOS}}-\rho c_{\mathrm{s}}^2\right)}{G c_{\mathrm{s}}^2}},
\end{equation}
where $ P_{\mathrm{EOS}} $ is the pressure from the Peng–Robinson equation in the present work, and it is defined as \cite{YuanPOF2006}
\begin{equation}
P_{\mathrm{EOS}}=\frac{\rho R T}{1-b \rho}-\frac{a \varphi(T) \rho^2}{1+2 b \rho-b^2 \rho^2}
\end{equation}
where $a=0.4572 R^2 T_c^2 / P_c, b=0.0778 R T_c / P_c$ and $\varphi=\left[1+\left(0.37464+1.54226 \omega-0.26992 \omega^2\right)\left(1-\sqrt{T / T_{\mathrm{c}}}\right)\right]^2$, in which $T_c$, $ p_c$ and $ \omega =0.344$ are the critical temperature, the critical pressure and the acentric factor, respectively. Here, the parameters $ a $, $ b $ and $ R $ are chosen as $ a=3/49 $, $ b=2/21 $ and $ R=1 $. The $T_{\mathrm{c}} \approx 0.10935$ can be calculated from the formulations of $ a $ and $ b $.

\subsection{The thermal LB model}

The new thermal LB model proposed by Wang et al. \cite{WangPRE2022} is adopted to solve the liquid-vapor phase change problem, and the most significant feature of this model compared with the previous models is that the Laplace term of the calculated temperature $[\nabla \cdot(\kappa \nabla T)]$ and the gradient term of the thermal capacitance $\nabla\left(\rho c_v\right)$ can be avoided. The evolution equation of the temperature distribution function can be described as
\begin{equation}
\rho c_v g_\alpha \left(\mathbf{x}+\mathbf{e}_\alpha \Delta t, t+\Delta t\right)-g_\alpha(\mathbf{x}, t)=\left(\rho c_v-1\right) g_\alpha\left(\mathbf{x}+\mathbf{e}_\alpha \Delta t, t\right)-\frac{1}{\tau_g}\left[g_\alpha(\mathbf{x}, t)-g_\alpha^{(e q)}(\mathbf{x}, t)\right]+\Delta t G_\alpha (1-s(\mathbf{x})),
\end{equation}
where $ g_\alpha $ is the temperature distribution function, $ c_v $ is the specific heat at constant volume, $ \tau_{g}= \lambda/c_s^{2}\Delta t+0.5 $ is the relaxation time, $ g_\alpha^{(e q)} $ is the the equilibrium distribution function for temperature defined as 
\begin{equation}
g_\alpha^{(e q)}=w_\alpha T \text {, }
\end{equation}
$ G_\alpha $ is the source term given by
\begin{equation}
G_i=-w_\alpha\left[\rho c_v \mathbf{u} \cdot \nabla T+T\left(\partial_T P_{E O S}\right)_\rho \nabla \cdot \mathbf{u}\right].
\end{equation}
The macroscopic temperature is calculated by the following equation
\begin{equation}
T=\sum_\alpha g_\alpha.
\end{equation}

\section{Problem statement}

	\begin{figure}[ht]
		\centering
		\includegraphics[width=0.6\textwidth]{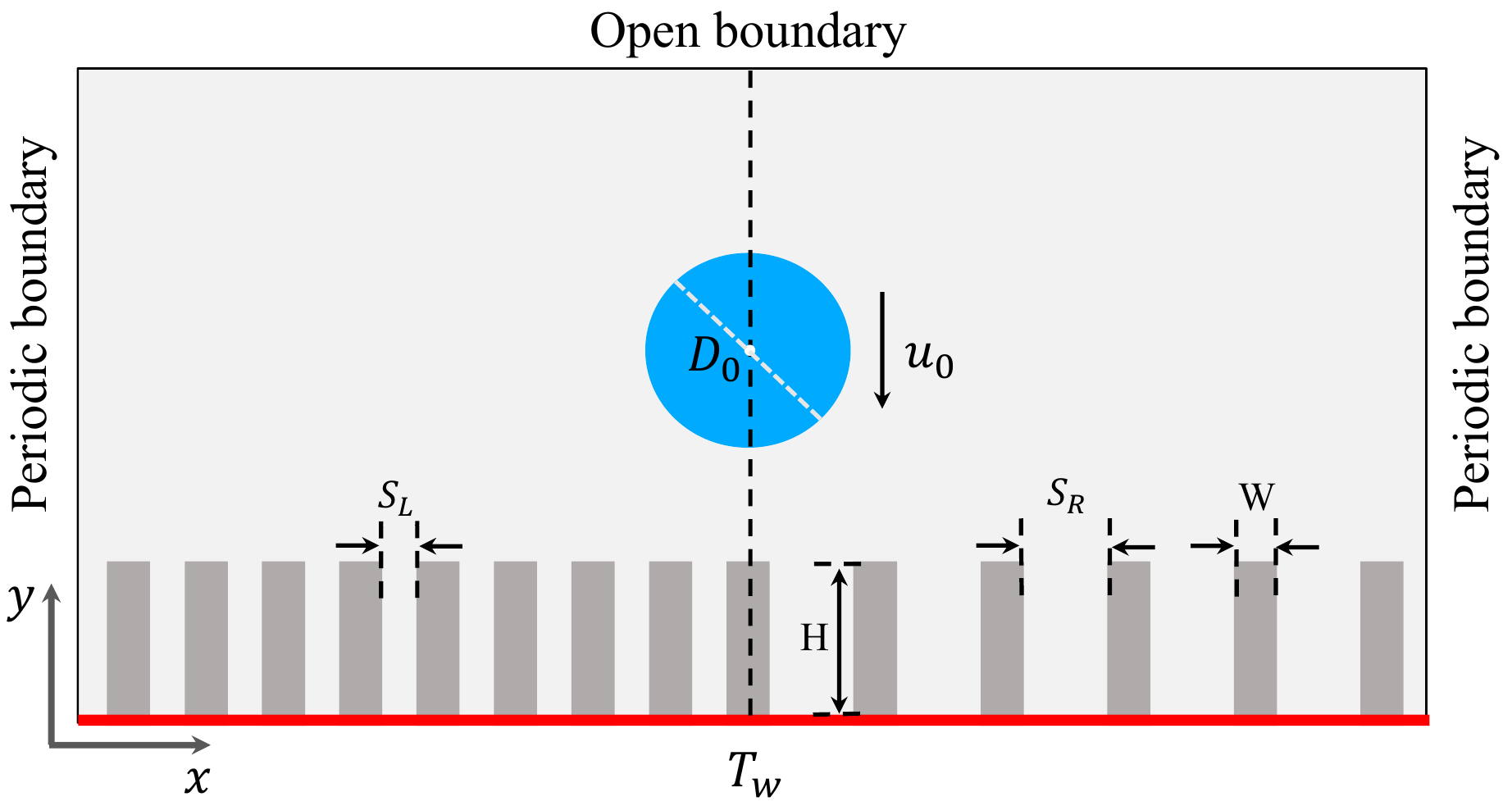}
		\caption{ Schematic illustration of droplet impact onto Janus-textured heated surface and the governing parameters.}
		\label{fig1}
	\end{figure}

Fig. \ref{fig1} shows the schematic of the droplet impacting the Janus-textured heated surface. A droplet with diameter $ D_0=40 $ and initial velocity $ u_0 $ is located above the center of the Janus-textured heated substrate. The Janus-textured substrate consists of uniformed micro arrays with ridge width $ W=2 lu $ and ridge height $ H $, while the groove width on both sides of the impingement center is different. The value of the pillar spacing $ S $ gradually increases from $ S_L=2lu $ in the denser region to $ S_R=10lu $ in the sparser region. In addition, the periodic boundary conditions are applied in the horizontal direction, the open boundary condition is employed in the top boundary and the bottom boundary is constant temperature boundary condition and no-slip boundary condition. In the simulation, the kinematic viscosity ratio is fixed at $ \nu_v/\nu_l=15 $, the saturation temperature is chosen to be $ T_{sat}=0.86T_c $, which corresponds to the liquid–vapor density ratio $ \rho_{l}/\rho_{v}=17.0 $. In addition, unless otherwise stated, the structural parameters of the surface are selected as above.

The process of droplet impacting on the heated Janus-textured surface is governed by the Weber number ($ We $), Jakob number ($ Ja $) and dimensionless time ($ t^* $), which are defined as
\begin{equation}
W e=\frac{\rho_l u_{0}^2 D_0}{\sigma}, \quad Ja=\frac{c_v\left(T_w-T_{s a t}\right)}{h_{lv}}, \quad t^*=\frac{t}{\sqrt{D_0 / g}},
\end{equation}
where $ \sigma $ is surface tension, $ T_{w} $ is the surface temperature, $ h_{lv}=0.5032 $ is the latent heat calculated by Ref. \cite{GongIJHMT2013}. The detailed simulation parameters are shown in Table 1. In addition, it is worth noting that all variables in present simulation is based on the lattice units.

\begin{table}[H]

	\caption{ Simulation parameters used in the present work.}
	\centering

    \begin{tabular}{cccc}
    \hline\hline
      								   & Liquid     & Vapor     & Solid            \\    \hline
         Density ($\rho$)              & 6.5        & 0.38      & 50.0             \\
    Specific heat ($c_v$)              & 8.0        & 4.0       & 10.0             \\
  Thermal conductivity ($\lambda$)     & 0.15       & 0.02      & 1.5              \\ \hline\hline
    \end{tabular}
    \label{table1}
\end{table}

\section{Model validation}

\begin{figure}[H]
		\begin{minipage}[t]{0.5\textwidth}
			\centering
			\includegraphics[width=\textwidth]{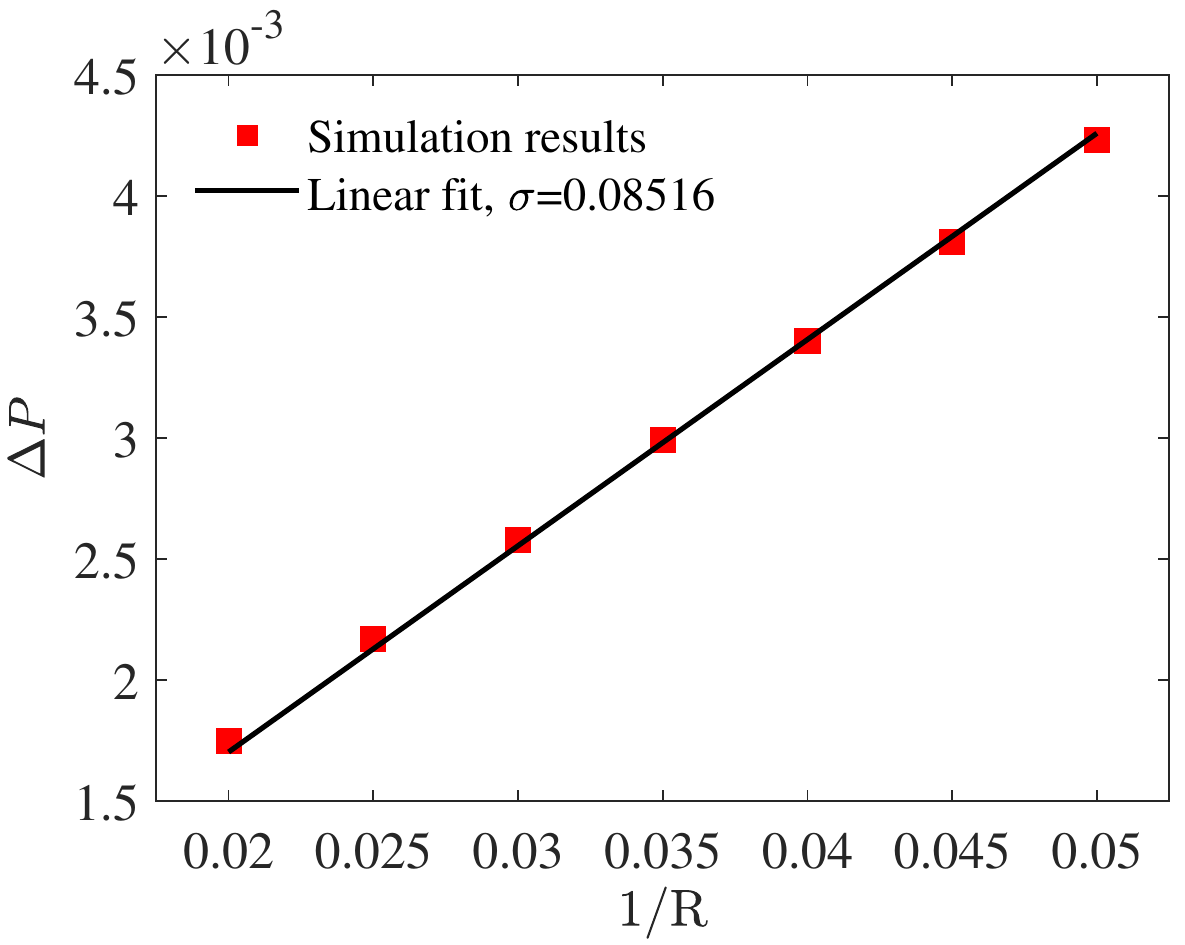}
			\caption{Evaluation of Laplace law at $ T/T_{c} = 0.86 $.}
			\label{fig2}
		\end{minipage}%
		\hfill
		\begin{minipage}[t]{0.5\textwidth}
			\centering
			\includegraphics[width=\textwidth]{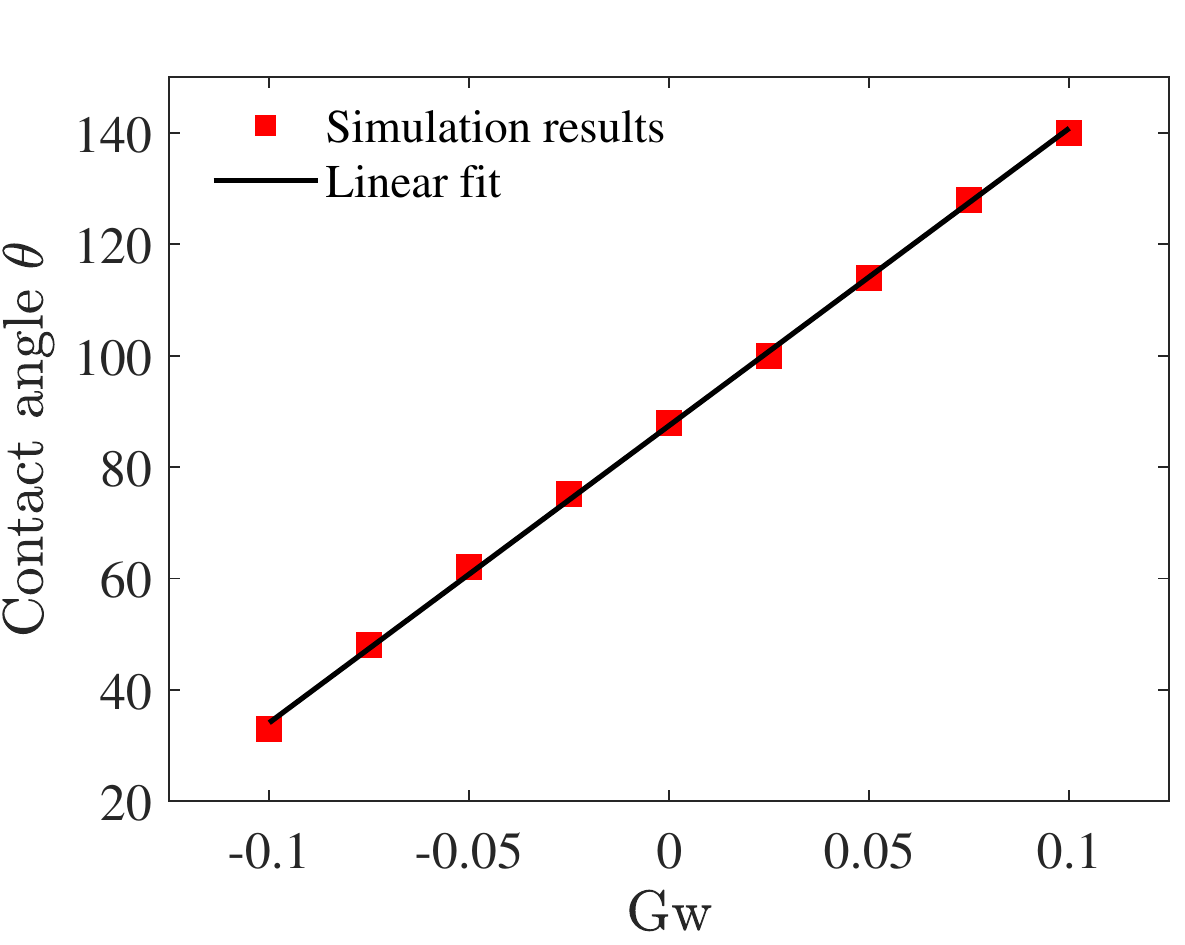}
			\caption{Contact angle $ \theta $ with respect to interaction strength Gw between the solid wall and fluid.}
			\label{fig3}
		\end{minipage}
	\end{figure}

In this section, three tests are considered to validate the present LB model. First, Laplace's law is adopted to verify the hydrodynamic effects of the present model. Laplace's law states that the pressure difference $ \Delta P $ inside and outside of the droplet is linear with the reciprocal of the droplet radius $ R $: $ \Delta P= \sigma /R $, where $ \sigma $ is the surface tension. In the simulation, a droplet placed at the center of the computational domain filled with saturated vapor. The periodic boundary conditions are applied to all boundaries. As shown in Fig. \ref{fig2}, the pressure difference is increase linearly with the $ 1/R $, which indicates the present LB model is consistent with Laplace's law and the surface tension can be calculated as 0.08516.

Secondly, we conduct the contact angle $ (CA) $ test by adjusting the $ G_w $ in Eq. \ref{eq9} to achieve different wettability. In our simulation, a droplet with $ R=50 lu $ is placed at the bottom center of $ Nx \times Ny = 400 \times 200 $ lattice system. The periodic boundary conditions are employed in the x-direction while the bounce-back scheme is applied for the rest of the boundaries. Fig. \ref{fig3} shows the measured contact angles for different $ G_w $ after the fluid system has reached an equilibrium state. We found that the contact angle is increased linearly with the increasing $ G_w $ and different $ G_w $ can achieve the adjustment of wettability of the surface.

	\begin{figure}[h]
		\centering
		\includegraphics[width=0.5\textwidth]{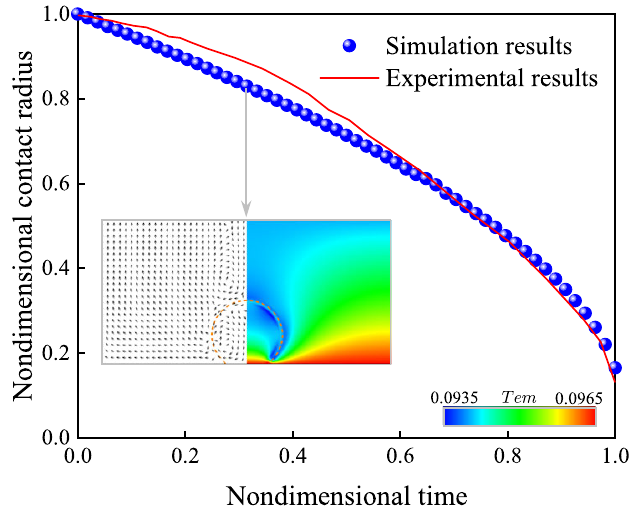}
		\caption{Comparisons between the present LB model simulations and experiments.}
		\label{fig4}
	\end{figure}
	
Finally, to validate the capacity of the present model for modeling the liquid-vapor phase change process, the droplet evaporation on the heated surface is simulated, and compare the numerical results against the experimental data by Dash et al. \cite{DashPRE2014}. In their experimental, a small with a size of $3 \pm 0.1 \mu 1$ is placed at the heated surface with an initial contact angle of $120^{\circ}$ and a temperature of $40^{\circ} \mathrm{C}$. The environmental temperature is $21 \pm 0.5^{\circ} \mathrm{C}$, and the gravity can be neglected due to the droplet being small. To maintain similarity with the experiment, we set the same Jacob number ($ Ja=0.036 $) in the simulation as in the experiment. Our simulations are conducted in a computation domain with $ Nx \times Ny = 400 \times 200 $, the droplet initially located in the bottom center of the heated surface surrounded by saturated vapor. As for the boundary conditions, the periodic boundary conditions are used in the x-direction, while the constant-pressure boundary condition is used for the top boundary. In addition, the temperature of the bottom and top boundaries are set to be $ T_w=0.8931T_c $ and $ T_s=0.86T_c $ in simulations. Fig. \ref{fig4} shows the evolution of nondimensional contact radius with the nondimensional time in the evaporation process and the temperature field distribution and vector distribution at $ t^*=0.29 $. We can observe a good agreement between our simulation results and experimental data, and the Marangoni flow that appears in the evaporated droplets also qualitatively validates our model.

\section{Results and discussions}

In what follows, the numerical simulation results of droplets impacting a Janus-textured heated surface are discussed and analyzed. The influences of the wettability, Jakob number, and Weber number are considered. To avoid the effect of some underlying factors, the setting of the other parameters is the same for all cases when we investigate the influence of one given factor.

\subsection{Wettability effect}

	\begin{figure}[H]
		\centering
		\includegraphics[width=1.0\textwidth]{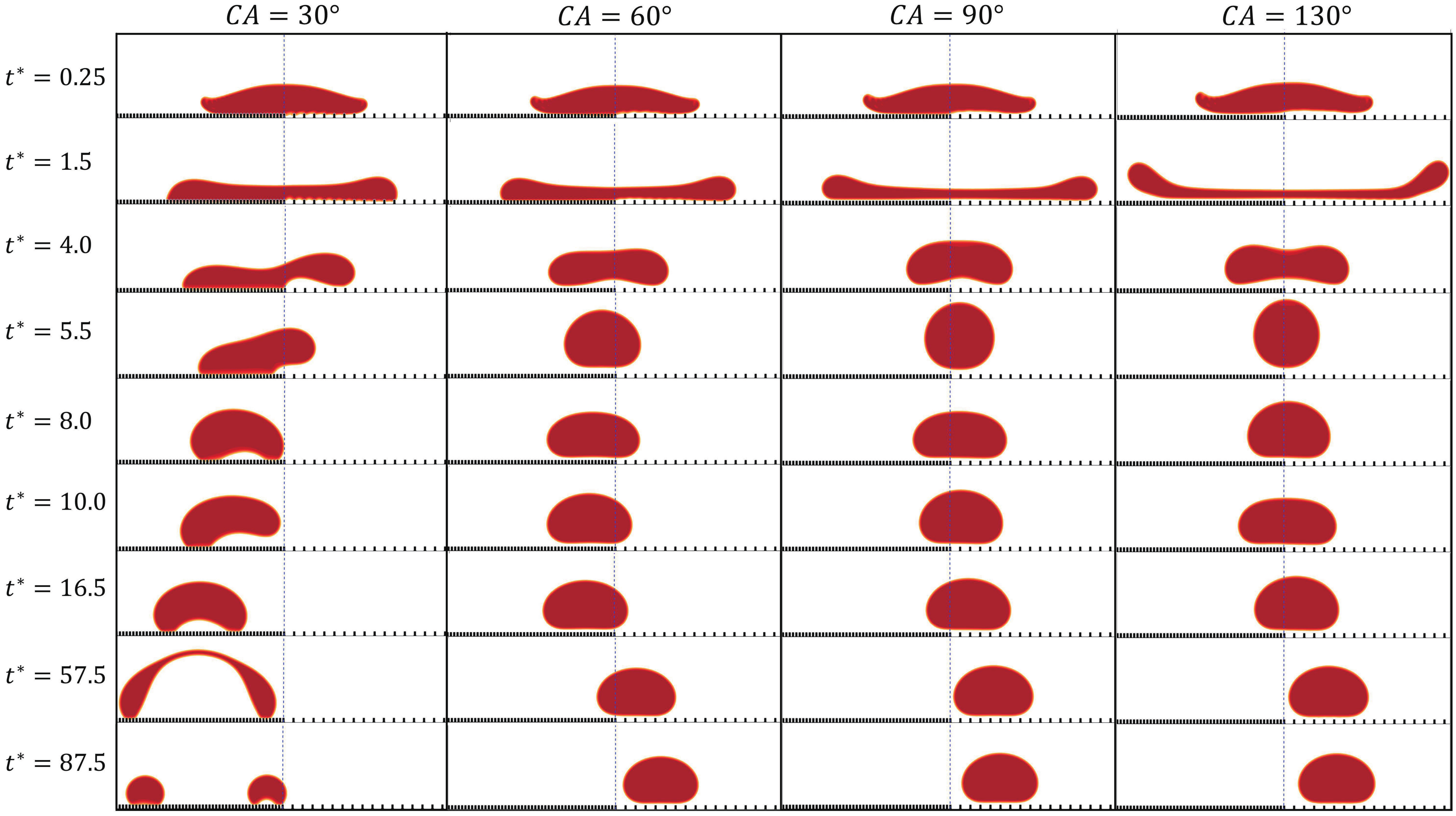}
		\caption{The morphologies of the droplets impinging upon the Janus-textured heated surface with different contact angles at $ We=122.10 $ and $ Ja=0.16 $.}
		\label{fig5}
	\end{figure}
	
According to the previous work on droplets impacting the solid substrate, the process of droplet impacting can be divided into the following four phases: the kinematic phase, the spreading phase, the relaxation phase, and the wetting or equilibrium phase \cite{HuangCF2022}. However, the evolutionary stages for droplets impacting the Janus-textured heated surface can be described by three phases based on the numerical results, i.e., the asymmetric spreading phase I, asymmetric retracting phase II and equilibrium evaporation phase III.

Fig. \ref{fig5} illustrates the evolution of the droplet impacting on the Janus-texted heated substrate with four different wettability of $ 30^{\circ} $, $ 60^{\circ} $, $ 90^{\circ} $ and $ 130^{\circ} $ at a constant $ We=122.10 $. For all cases from $ t^*=0.25 $ to $ t^*=1.5 $, the droplet spreads rapidly under the action of the initial kinetic energy until the maximum spread diameter is reached when all initial impact energy is converted to surface energy, in which the case of $ CA=130^{\circ} $ has maximum spreading length due to low viscous dissipation for the low wettability. For the large wettability at $CA= 30^{\circ} $, the vapor bubbles generated among the sparser Janus-textured substrate coalesce to form a vapor pocket. With the growth of the vapor pocket, the droplet gradually migrates directionally on the surface toward the denser region. Then the vapor bubble nucleates and grow within the droplet. Finally, the droplet split into two sub-droplets due to the bursting of the growing vapor bubble. This case is also named transition boiling in some previous studies \cite{XuATE2022,WangPOF2022}. However, for the relatively low wettability at $CA= 60^{\circ} $, $CA= 90^{\circ} $ and $CA= 130^{\circ} $, the directional rebound behavior is occur due to the unbalanced interfacial forces created by the heterogeneous architectures. Subsequently, the vapor layer underneath the droplet is generated, which keeps the droplet levitated on the Janus-textured surface. Finally, the droplet gradually moves toward the sparser side from the impingement center due to the asymmetric vapor pressure caused by asymmetric structures. In addition, the Janus-textured surface with low wettability has superior nucleation behavior resulting in their typically easy transition from efficient transition boiling to inefficient film boiling.

	\begin{figure}[H]
		\centering
		\includegraphics[width=0.6\textwidth]{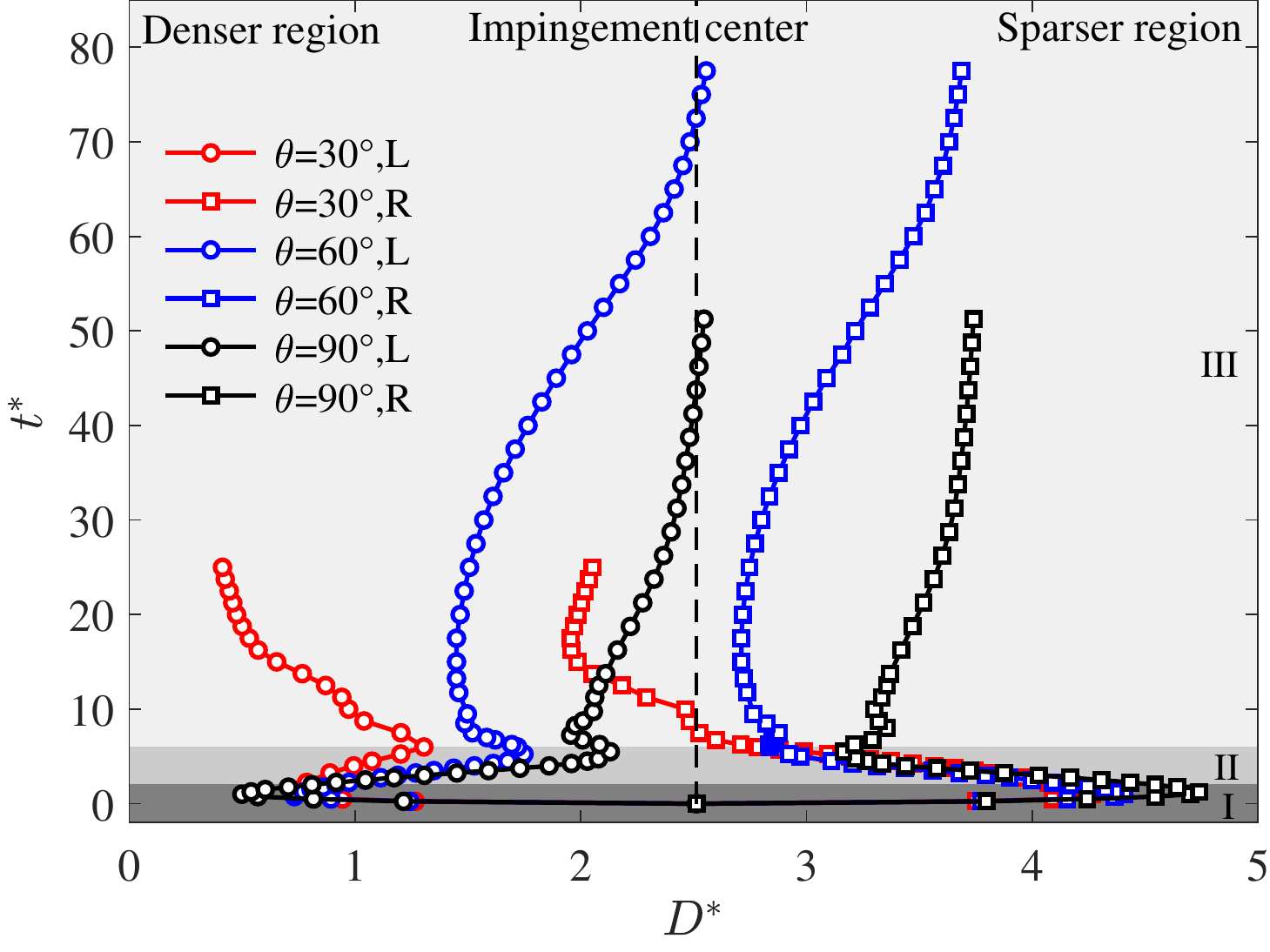}
		\caption{ Temporal evolution of the normalized maximum spread diameter for different wettability differences at $ We=122.10 $ and $ Ja=0.16 $.}
		\label{fig6}
	\end{figure}


Fig. \ref{fig6} shows the temporal evolution of the normalized maximum spread diameter $ D^* $ for different wettability, in which the spreading diameter is defined as the maximum distance of the droplet in the horizontal direction. In the asymmetric spreading phase I, the spread factor is large in the sparser region due to the low wettability of the sparser region, and the maximum $ D^* $ increases lightly with the increasing contact angle. Following the initial asymmetric spreading phase, the droplet retracts rapidly toward the impingement center during the asymmetric retracting phase II. The droplet detaches itself from the sparser region and completely migrates to the denser region for the case of $CA= 30^{\circ} $. Finally, in the equilibrium evaporation phase III, the droplet moves rapidly toward the sparser region at the film boiling state for the case $CA= 60^{\circ} $ and $CA= 90^{\circ} $. In conclusion, the simulation results show that the droplets eventually migrate toward the denser region at the transition boiling state, while the droplets eventually migrate toward the sparser region at the film boiling state.

\begin{figure}[H]
		\begin{minipage}[t]{0.5\textwidth}
			\centering
			\includegraphics[width=\textwidth]{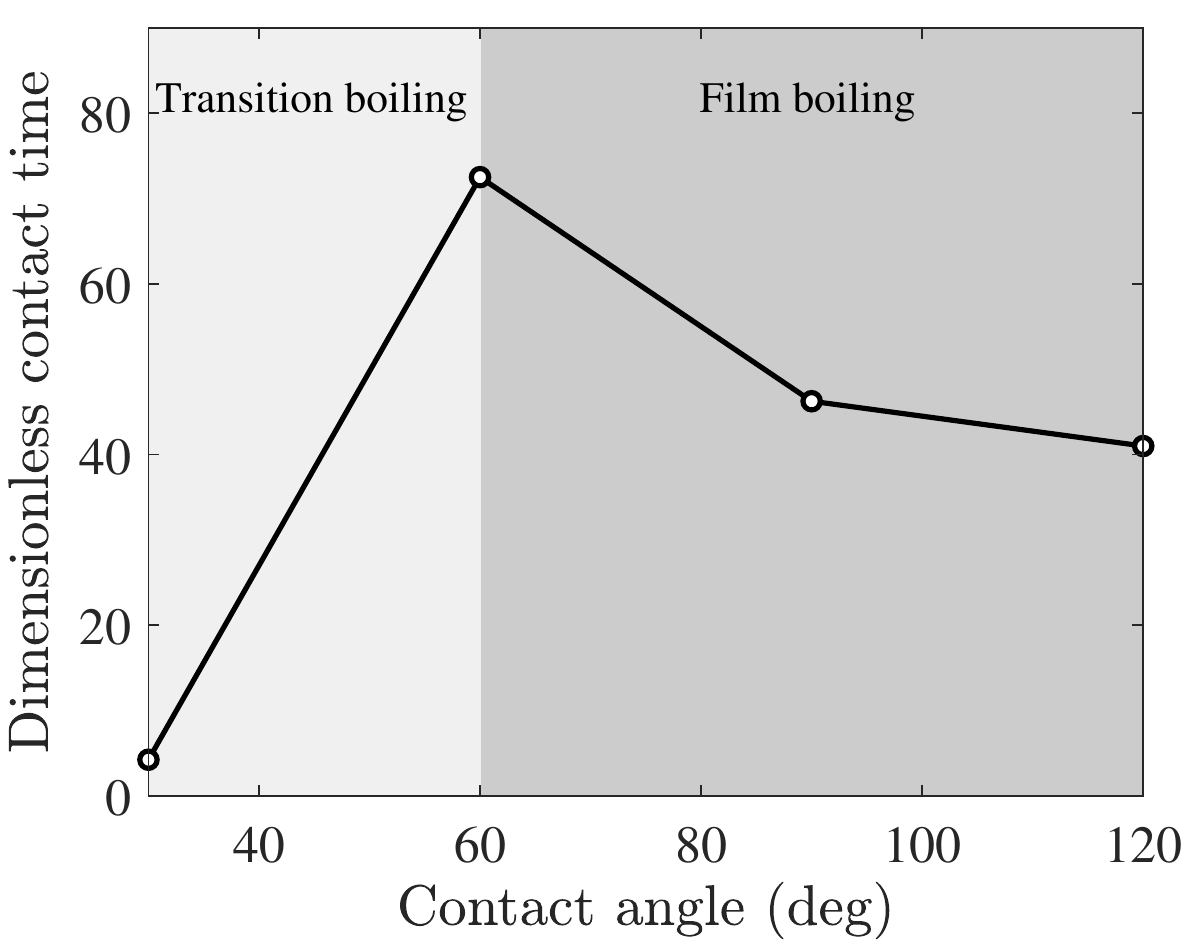}
			\caption{Variation of dimensionless contact time with different contact angles.}
			\label{fig7}
		\end{minipage}%
		\hfill
		\begin{minipage}[t]{0.5\textwidth}
			\centering
			\includegraphics[width=\textwidth]{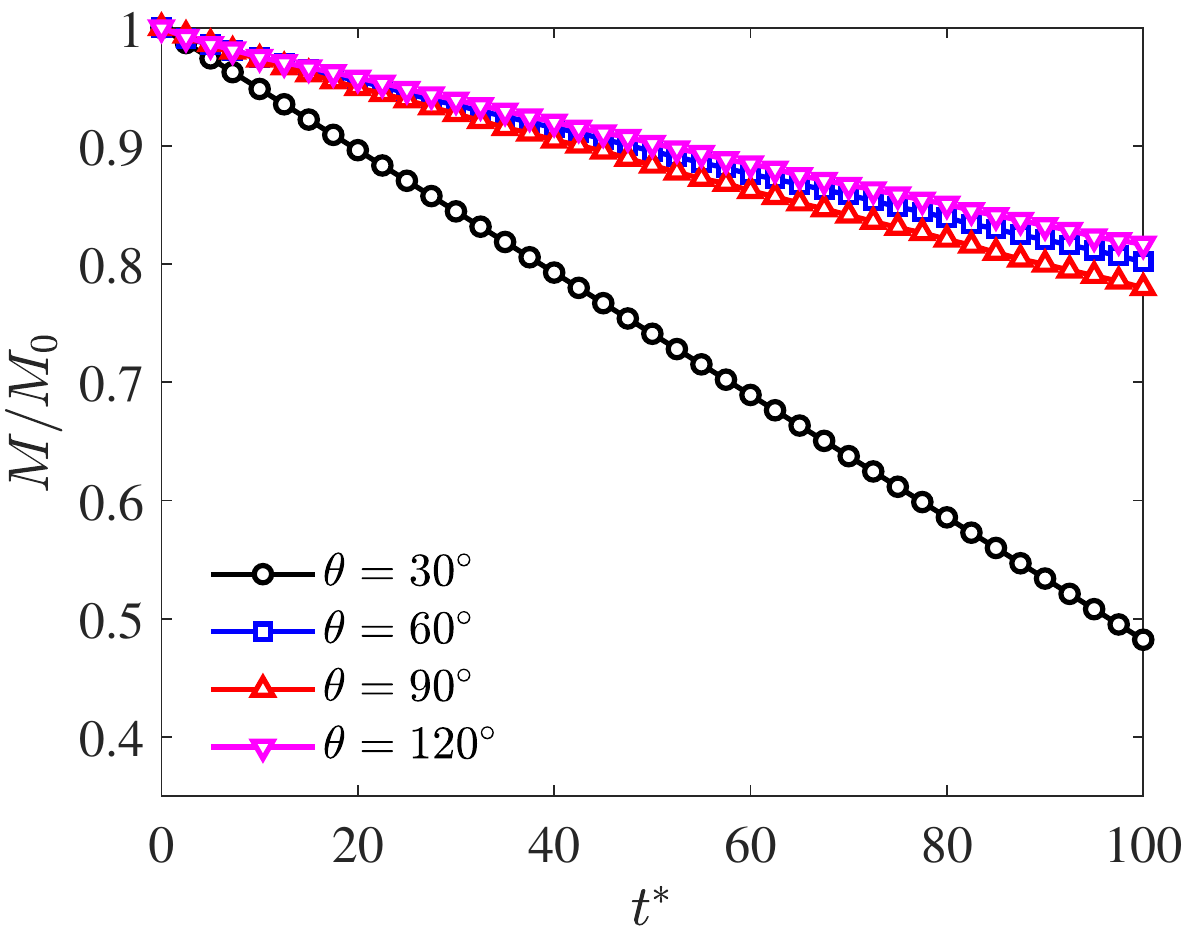}
			\caption{Time evolution of the dimensionless droplet mass for different contact angles.}
			\label{fig8}
		\end{minipage}
	\end{figure}

The contact time, defined as the duration of the droplet completely migrating to one side of the surface, is shown in Fig. \ref{fig7} at different contact angles. It can be seen that the contact time of droplets in the transition boiling state is shorter than that in the film state, which is caused by the fact that the different boiling states influence the dynamic of the impacting droplet. For the transition boiling state at $CA= 30^{\circ} $, the droplet completely shrank from the sparser region to the denser region under the action of the surface tension force and the growing vapor bubbles in the asymmetric retracting phase II. However, for the film boiling at $CA= 60^{\circ} $, $CA= 90^{\circ} $ and $CA= 130^{\circ} $, a part of the droplet remains in the denser region after asymmetric retracting phase II, which migrates slowly toward the sparse region under the action of the vapor film and vapor flow. In addition, the contact time decreases with increasing contact angle for the film boiling, which is mainly due to the fact that the droplets first contact the surface with less energy dissipation for larger contact angles.

Fig. \ref{fig8} illustrates the variation of the dimensionless droplet mass with time for different contact angles. It is apparent that the droplets in the transition boiling state evaporate more rapidly compared to the droplets in the film boiling state, which is consistent with experiments [*]. For the transition boiling regime at $CA= 30^{\circ} $, the vapor bubble behavior seriously distorts the droplet interface during the vigorous boiling heat transfer, which greatly reduces the evaporation time of the droplet. However, for the film boiling regime at $CA= 60^{\circ} $, $CA= 90^{\circ} $ and $CA= 130^{\circ} $, a vapor layer is generated between the droplet and the Janus-textured surface, which considerably reduces the heat transfer and therefore retards the evaporation of the droplet.

\subsection{Jakob number effect}

	\begin{figure}[H]
		\centering
		\includegraphics[width=1.0\textwidth]{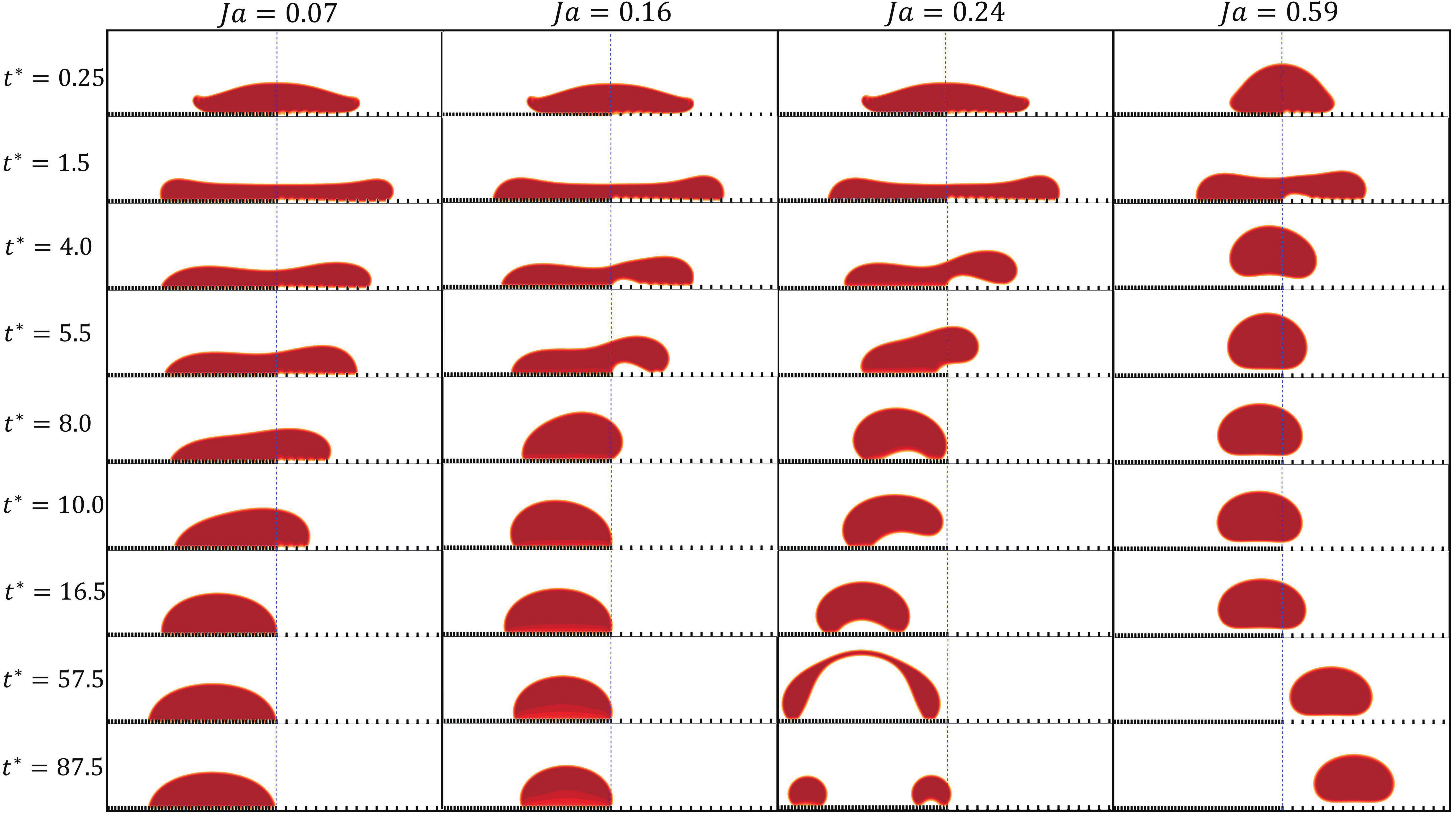}
		\caption{The morphologies of the droplets impinging upon the Janus-textured heated surface with different contact angle at $ We=122.10 $ and $CA= 30^{\circ} $.}
		\label{fig9}
	\end{figure}

We next consider the influence of the Jakob number on the boiling state of the impacting droplet on the Janus-textured heated surface. In our simulation, the variation of the Jakob number is obtained by adjusting the temperature of the surface. Fig. \ref{fig9} shows the whole process of droplet impingement on the Janus-textured heated surface for different Jakob numbers, which include three different boiling states, i.e., contact boiling, transition boiling and film boiling. For all cases from $ t^*=0.25 $ to $ t^*=1.5 $, the droplet penetrates the groove space in the denser region and shows a partial Wenzel state, while the other side remains in the partial Cassie state as a result of the growing vapor pocket. The different wetting states result in different contact angles on both sides of the impingement center, which generate an unbalanced Young's force. For relatively low superheat ($ Ja=0.07 $), the droplet eventually migrates to the side of the denser region under the unbalanced Young's force, with no vapor bubble formation during the migration process. With an increase in surface superheat ($ Ja=0.16 $), the right side of the droplet leaves the surface since the coalescence of the growing vapor bubble within the droplet. Subsequently, the droplet evaporates on the sparser region without forming the vapor bubble, which is called contact boiling. As the surface superheat increases further ($ Ja=0.24 $), the droplet in the transition boiling state splits into two sub-droplets due to the bursting bubble after migration to the denser region. When surface overheating is relatively high  ($ Ja=0.59 $), the droplets rebound directionally toward the side of the denser region in the asymmetric retracting phase II, then a stable vapor layer is formed between the droplet and the Janus-textured surface. The droplet eventually migrates to the sparser region due to the asymmetry of the structure and the vapor flows beneath the droplets.

\begin{figure}[H]
		\centering
		\includegraphics[width=0.5\textwidth]{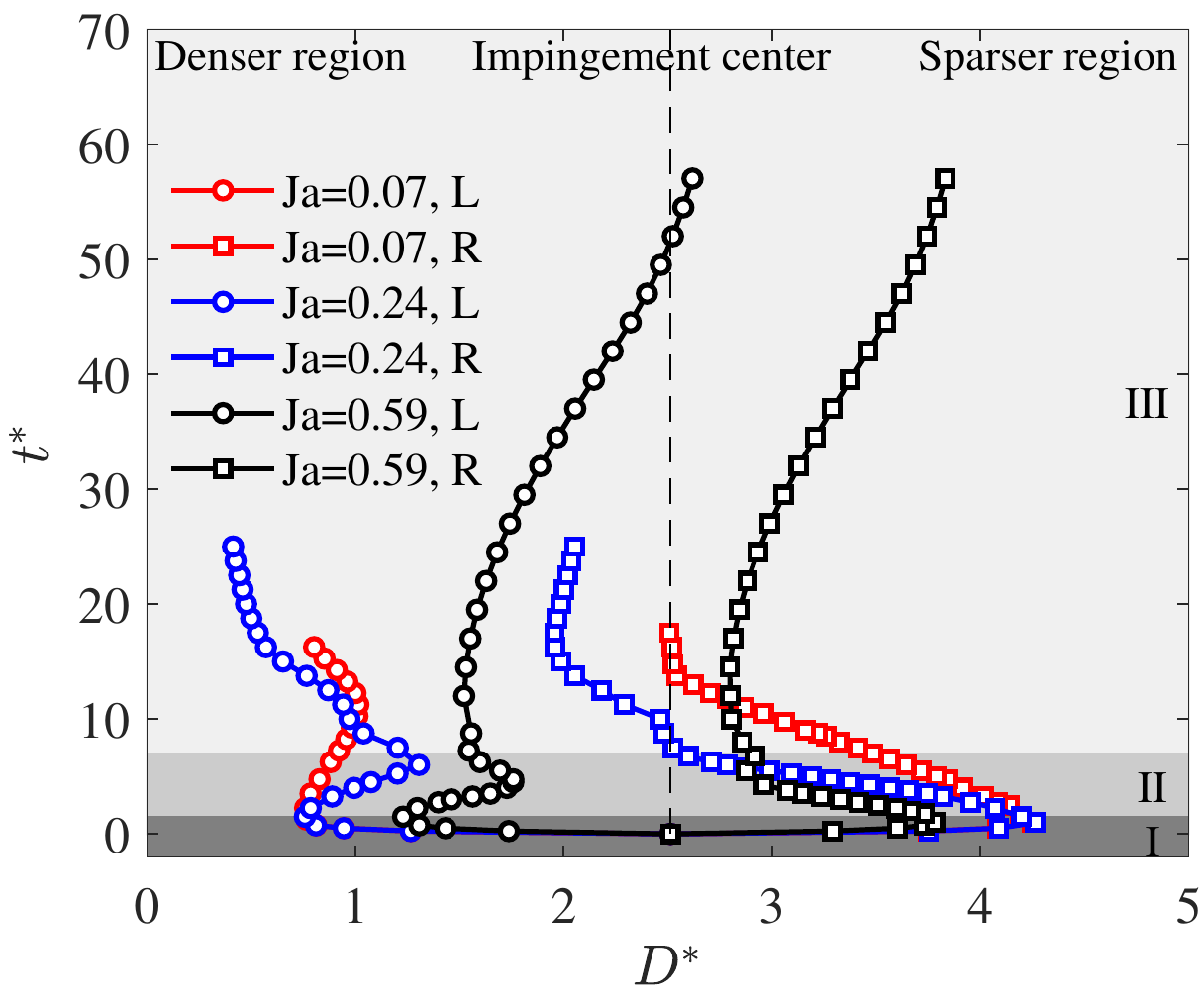}
		\caption{ Temporal evolution of the normalized maximum spread diameter for different wettability differences at $ We=122.10 $ and $CA= 30^{\circ} $.}
		\label{fig10}
	\end{figure}

\begin{figure}[H]
		\begin{minipage}[t]{0.5\textwidth}
			\centering
			\includegraphics[width=\textwidth]{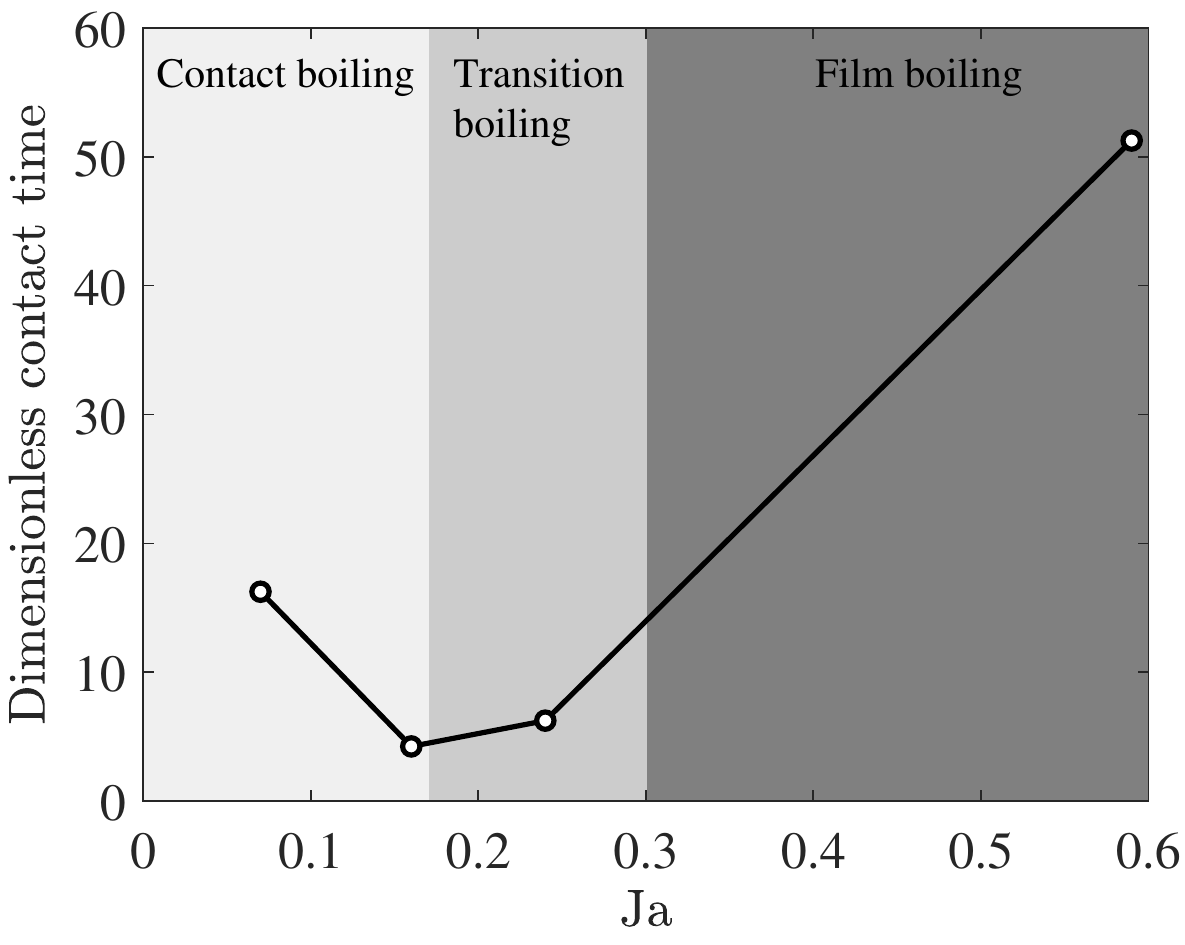}
			\caption{Variation of dimensionless contact time with different Jakob numbers.}
			\label{fig11}
		\end{minipage}%
		\hfill
		\begin{minipage}[t]{0.5\textwidth}
			\centering
			\includegraphics[width=\textwidth]{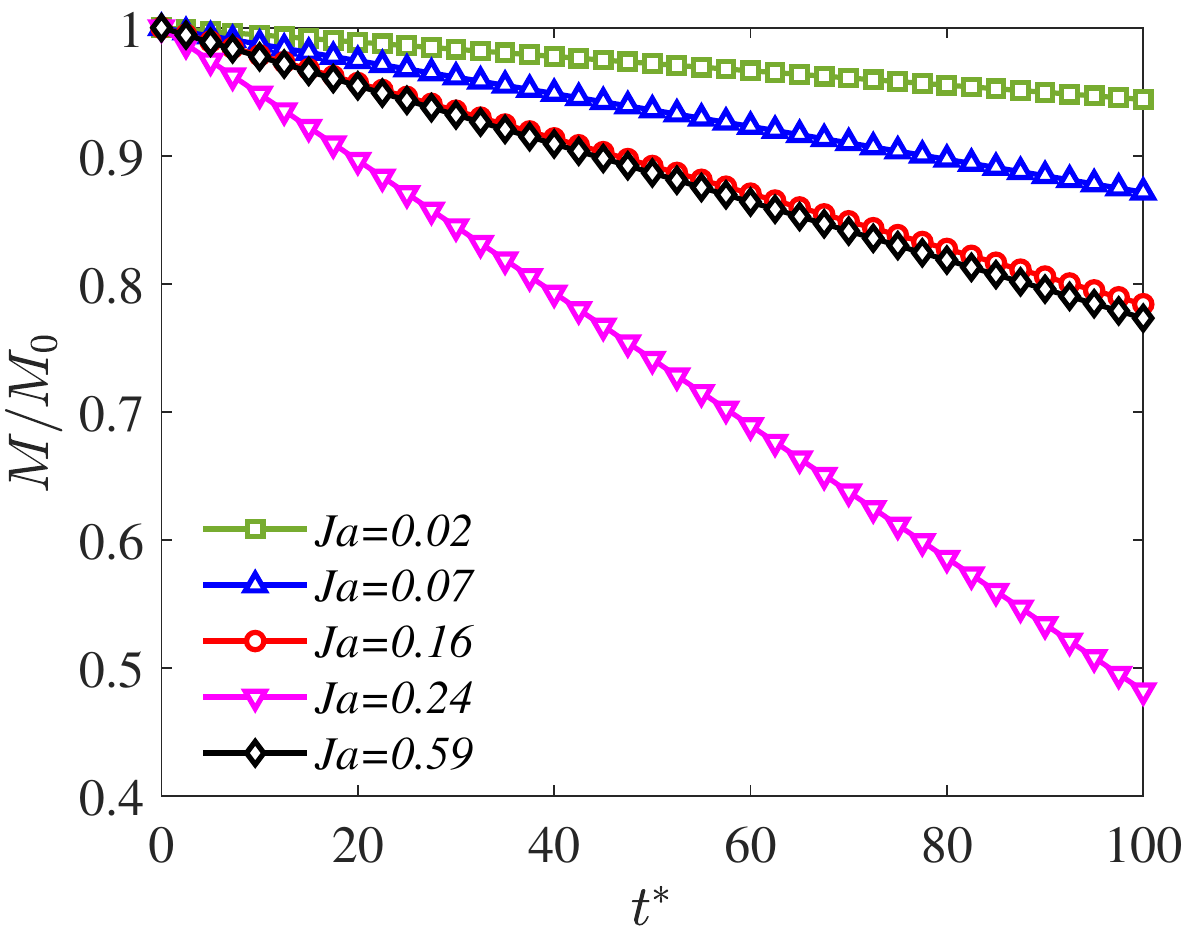}
			\caption{Time evolution of the dimensionless droplet mass for different Jakob numbers.}
			\label{fig12}
		\end{minipage}
	\end{figure}

To have a deeper insight on the droplet dynamic behavior in this case, Fig. \ref{fig10} shows the evolution of the spreading factor with dimensionless time for three different boiling regimes. It can be seen that in the asymmetric spreading stage I, the spreading factor increases with the decrease of Jakob number. For the contact and transition boiling states, the droplet tends to migrate the denser microstructure region, while the droplet eventually located  on the sparser region for the film boiling regime. Fig. \ref{fig11} plots the variation of contact time at different Jakob numbers. The contact time of droplet in contact boiling and transition boiling is shorter than that of droplet in film boiling regime. Fig. \ref{fig12} shows the variation of the dimensionless droplet mass with time for different Jakob numbers, it can be seen that the droplet evaporation rate increases with increasing Jakob number. For the relatively low surface temperature ($Ja=0.02$, $Ja=0.07$), the droplets remain in contact with the Janus-textured surface until the droplet completely evaporates due to the fact that the heat flux is insufficient to trigger bubble nucleation. There is no vapor bubble nucleation in the droplets, which leads to poor heat transfer performance. With the surface temperature increase, the droplet changes from contact boiling state to transition boiling state, the droplet interface is severely deformed by the evaporation and  nucleation process, which greatly enhances the nucleate boiling heat transfer performance, and the droplet evaporation is fast. However, the droplets show the film boiling state as the Jakob number further increases, and the evaporation rate of the droplets decreases significantly due to the fact that the stable vapor layer as a thermal insulation layer drastically prevents the heat transfer between the droplet and the Janus-textured heated surface.

\begin{figure}[H]
		\centering
		\includegraphics[width=0.7\textwidth]{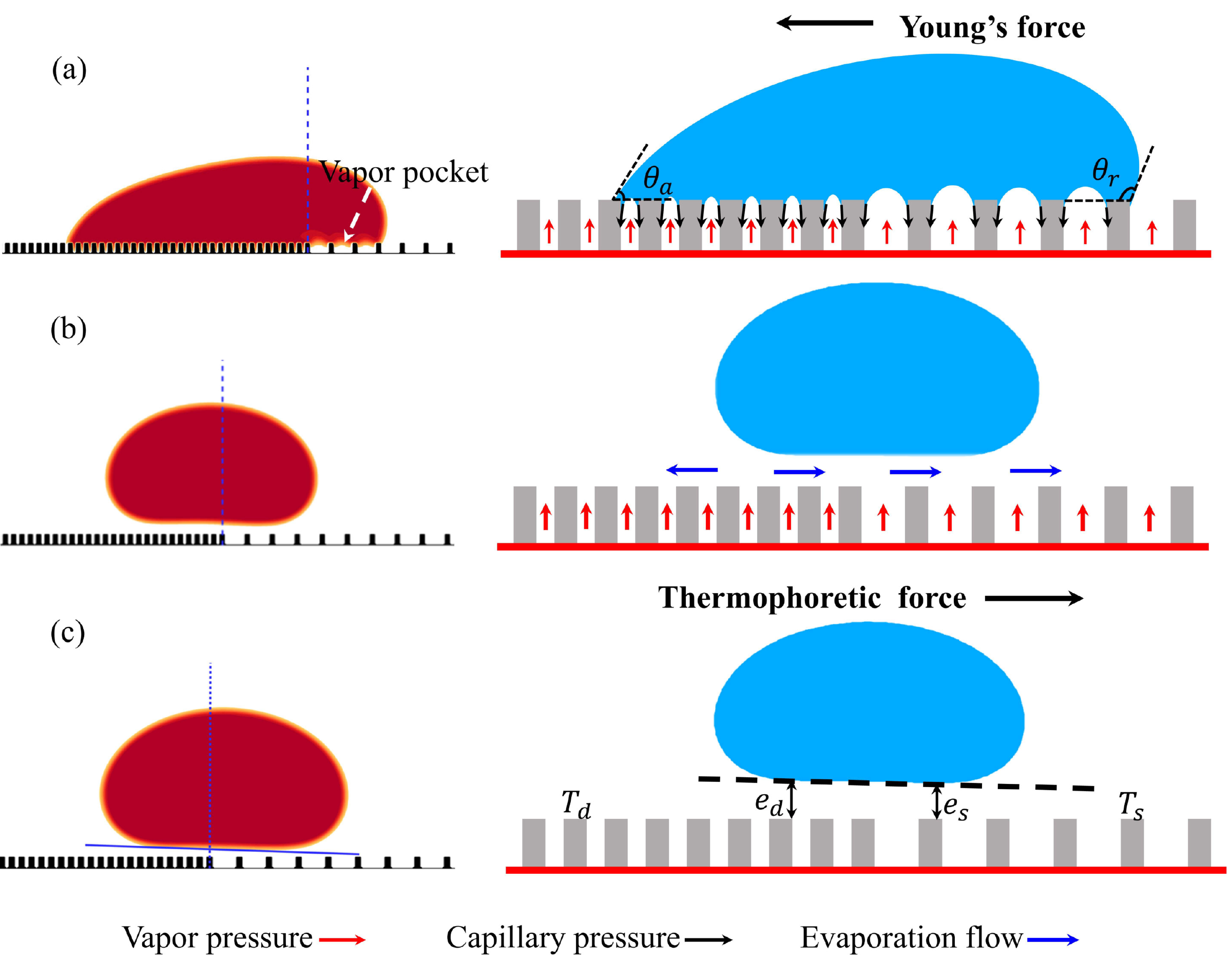}
		\caption{Mechanisms of different droplet motion behaviors on Janus-textured substrates}
		\label{fig13}
	\end{figure}

We next discuss the directional transport mechanism of droplets on the heated Janus-textured surface. The contact angles of droplets deposited on the surfaces with uniformed micropillars at saturation temperature are obtained by \cite{PatankarLangmuir2010}
\begin{equation}
\cos \overline{\theta}=r \cos \theta=\left(1+\frac{2 h}{w+s}\right) \cos \theta, 
\end{equation}
where $ r $ is the roughness of the microstructured surface, $ w $ is the ridge width, $ h $ is the ridge height, $ s $ is the pillar spacing, $\theta$ is the intrinsic contact angles, $\overline{\theta}$ is the apparent contact angles. It follows that the apparent contact angle of the Janus textured surface gradually increases from $ \overline{\theta}_{l} $ in the denser region to $ \overline{\theta}_{r} $ in the sparser region due to the variation in pillar spacing, resulting in a difference in wettability between both sides of the Janus-textured surface. Moreover, the characteristic curvature of the interface causes a downward capillary pressure \cite{KwonAPL2013,BartoloEPL2006,KwonPRL2011}
\begin{equation}
P_{\mathrm{cap}}=\frac{\sigma(\pi w) \cos \theta}{(w+s)^2}.
\end{equation}
It is clear that the capillary increases with the decreasing pillar spacing $ s $. Therefore, a larger capillary force is created at the denser region, which hinders the movement of the contact line, resulting in a smaller receding contact angle compared with that in the sparser region  [see in Fig. \ref{fig13}(a)]. In conclusion, an unbalanced force on the opposite sides of the impingement center is generated \cite{ChaudhuryScience1992,MalouinAPL2010}
\begin{equation}
d F_u= \gamma\left(\cos \theta_{\mathrm{r}}-\cos \theta_{\mathrm{a}}\right) dl,
\end{equation}
where $ \gamma $ is the liquid-vapor surface tension, $ \theta_{\mathrm{r}} $ is the receding contact angle, $ \theta_{\mathrm{a}} $ is the advancing contact angle, and $ l $ is the length of the contact line. Based on the Eq. (21*), we can conclude that the droplet generates an unbalanced Young's force toward the high-wettability side for $\theta_{\mathrm{a}}<\theta_{\mathrm{r}}$, while the droplet generates an unbalanced Young's force toward the low-wettability side for $\theta_{\mathrm{a}}>\theta_{\mathrm{r}}$. When a droplet retracts on the Janus-textured surface at contact boiling state, there is an unbalanced Young force towards the denser region due to $\theta_{\mathrm{a}}<\theta_{\mathrm{r}}$, which causes the droplet to move toward the denser region. In addition, a concave meniscus is generated by the downward capillary pressure and the upward vapor pressure at the liquid-solid interface. 

For the film boiling regime, the continuous vapor layer separating the droplet from the Janus-textured heated surface eliminates the unbalanced Young's force [see in Fig. \ref{fig13}(b)]. Therefore, the dynamics of the droplet in the film boiling state are determined by the vapor pressure. The vapor pressure under the droplet can be approximated by modeling the vapor flow as a viscous-dominated radial Poiseuille flow \cite{AvedisianIJHMT1987,BiancePOF2003}. The shear loss of vapor film has two components: one is caused by the velocity gradient of the vapor flow above on the surface and scales as $ \mu_{\mathrm{v}} V_v/e $, where $ \mu_{\mathrm{v}} $ is the dynamic viscosity, $ e\approx h $ is the thickness of the vapor layer and $ V_v $ is the velocity of vapor. The other is caused by the surface structure and scales as $ \mu_{\mathrm{v}} V_v/s $. Therefore, the total pressure formed in the vapor layer is computed by the above two parts and can be expressed as \cite{KwonAPL2013}
\begin{equation}
P_{\mathrm{vap}} \approx \frac{\mu_{\mathrm{v}} q \lambda^2}{\rho_{\mathrm{v}} h_{\mathrm{lv}} h^2 K},
\end{equation}
where $ \lambda $ is the length of the contact patch, $K \approx\left(1 / e^2+1 / s^2\right)^{-1}$ is the approximate permeability. The conductive heat flux $q$ can be expressed $ q= k_{\mathrm{eff}}(T_{\mathrm{w}}-T_{\mathrm{sat}}) $, where $ k_{\mathrm{eff}} $ is the effective heat conductivity. The effective heat conductivity $ k_{\mathrm{eff}} $  depends on the surface geometry and can be obtained by 
\begin{equation}
k_{\mathrm{eff}}=k_v+\left(k_s-k_v\right) /(1+w / s)^2,
\end{equation}
where $ k_v $ and $ k_s $ represent the thermal conductivity of the vapor and the solid surface, respectively. The effective heat conductivity $ k_{\mathrm{eff}} $ decreases as the pillar spacing $ s $ increases, it indicates that the sparser region with smaller $ k_{\mathrm{eff}} $ weakens the heat flux $ q $. Combining Eqs. (22) and (23) yield a general expression for the vapor pressure
\begin{equation}
P_{\mathrm{vap}} \approx\left(\frac{1}{h^2}+\frac{1}{s^2}\right) \frac{\mu_{\mathrm{v}} k_{\mathrm{eff}}\left(T_{\mathrm{s}}-T_{\mathrm{sat}}\right) \lambda^2}{\rho_{\mathrm{v}} h_{\mathrm{lv}} h^2}.
\end{equation}
According to Eq (24), with the increase of the pillar spacing $ s $ from the denser region to the sparser region, the effective thermal conductivity $ k_{\mathrm{eff}} $ decreases, resulting in the vapor pressure $ P_{\mathrm{vap}} $ in the sparse region decreases. In conclusion, the pressure difference developed in the vapor bubble/film on two sides of the Janus-textured surface due to the spatial gradient in the conductive heat flux, which promotes the droplet directionally transport to the sparser region. 

In addition, according to the Eq. (23*), it can be seen that the effective heat conductivity $ k_{\mathrm{eff}} $ in denser region is larger than that in sparser region, which will lead to inhomogeneous temperature on both sides fo the impingement center. As shown in Fig. \ref{fig13}(c), the inhomogeneous temperature can incline the drop base and induce propulsion toward sparser region since the evaporation of the droplet depends on the substrates temperature \cite{SobacPOF2017,BouillantSM2021}.

\subsection{Weber number effect}

	\begin{figure}[H]
		\centering
		\includegraphics[width=1.0\textwidth]{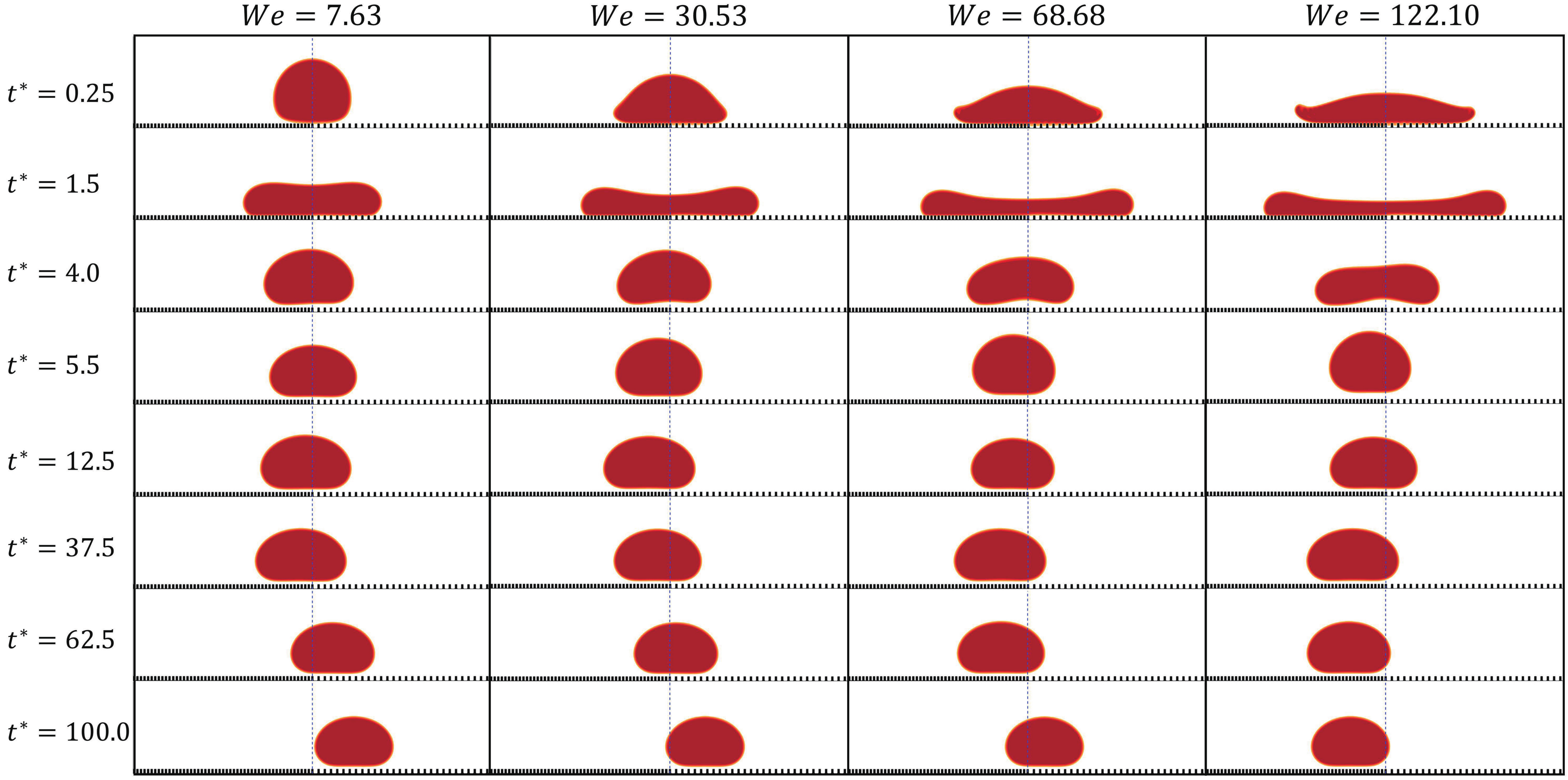}
		\caption{The morphologies of the droplets impinging upon the Janus-textured heated surface with different Weber numbers at $ S_{r}=5lu $, $CA= 60^{\circ} $ and $ Ja=0.24 $.}
		\label{fig14}
	\end{figure}

Now attention turns to the influences of the Weber number on the directional transport of the droplet on the Janus-textured surface. In our simulations, the Weber number is adjusted by the initial velocity $ u_0 $, resulting in the kinetic energy of the droplet increasing with the increasing Weber number. Fig. \ref{fig14} illustrates the evolution of the impinging droplet on the Janus-textured heated surface for different Weber numbers. It can be seen from fig. 14[]* that with the Weber number increases, the droplet with more kinetic energy has a large spreading length, which increases the spreading velocity of the droplet and the contraction velocity after impacting the surface. For all cases, the droplet bounces directionally toward the denser region due to the unbalanced Young's force toward the denser region after it spreads to the maximum spreading length. Then the vapor layer is generated between the droplet and the Janus-textured surface, resulting in the droplet shows in the film boiling state. For the low Weber numbers, the droplets eventually migrate toward the sparser region since the fact that vapor pressure difference generated by the different structures on the Janus-textured surface as a driving force to propel the Leidenfrost droplet directionally, while the droplets remain in the denser region for large Weber number. 

	\begin{figure}[H]
		\centering
		\includegraphics[width=0.6\textwidth]{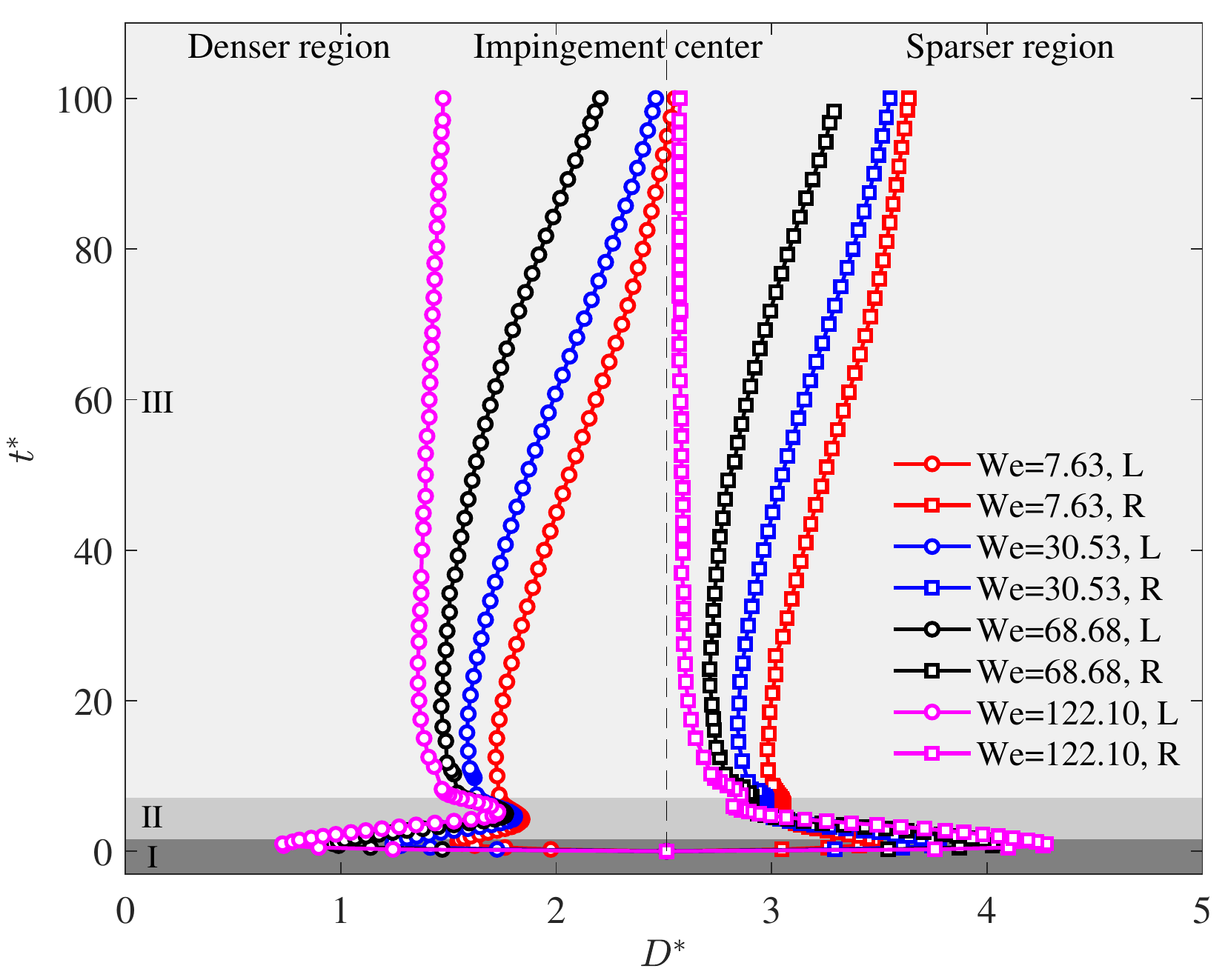}
		\caption{Temporal evolution of the normalized maximum spread diameter for different wettability differences at $ S_{r}=5lu $, $CA= 60^{\circ} $ and $ Ja=0.24 $.}
		\label{fig15}
	\end{figure}

To further understand the effects of Weber number on the behaviors of impinging droplets, Fig. \ref{fig15} illustrates the temporal evolution of the spread factor $ D^* $ for different Weber numbers. As shown in this figure, it is apparent that the maximum spreading factor increases with the Weber number increase due to more kinetic energy in the asymmetric spreading phase I. After the asymmetric retracting phase II, the droplets gradually transport directionally to the sparser region under the effect of the vapor pressure difference for $ We=7.63 $, $ We=30.53 $ and $We=68.68$. As the Weber number increases to $ We=122.10 $, the droplet with more initial energy bounces completely to the denser region due to the large unbalanced Young's force after the initial contact with the surface. Therefore, the droplet without the effect of vapor pressure is located in the denser region. 

	\begin{figure}[H]
		\centering
		\includegraphics[width=0.9\textwidth]{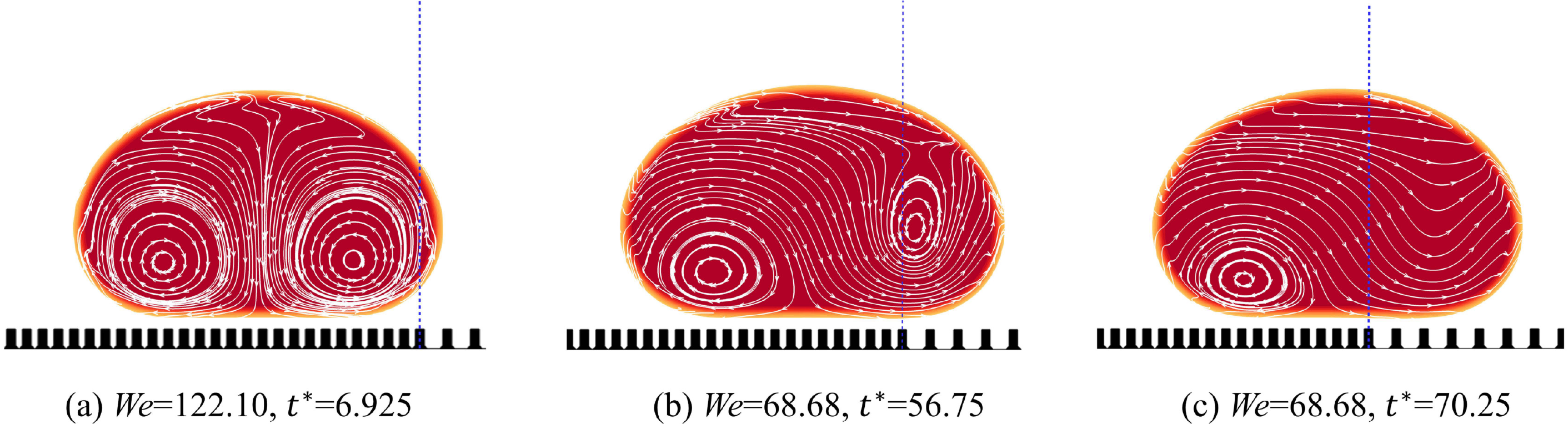}
		\caption{The streamlines of the case $ We=122.10 $ and $ We=68.68 $ at different times.}
		\label{fig16}
	\end{figure}

To further illustrates the effect of vapor pressure difference on the directional transport behavior of droplet, Fig. \ref{fig16} displays the streamlines inside the droplet of the case $ We=122.10 $ and $ We=68.68 $ at different times. For the case of $ We=122.10 $ at $ t^*=6.925 $, the droplets without the influences of the vapor difference completely remained in the denser region, and two approximately symmetrical vortices that rotate in opposite directions are formed inside the droplet due to the Marangoni effect. However, for the case of $ We=68.68 $, the droplets driven by the vapor pressure gradient gradually migrate toward the sparser region. During the migration phase, the counter-clockwise vortex gradually disappears due to the fact that the droplet and the vapor move in the same direction on the right side. Meanwhile, the clockwise vortex has become very small.



	\begin{figure}[H]
		\centering
		\includegraphics[width=0.6\textwidth]{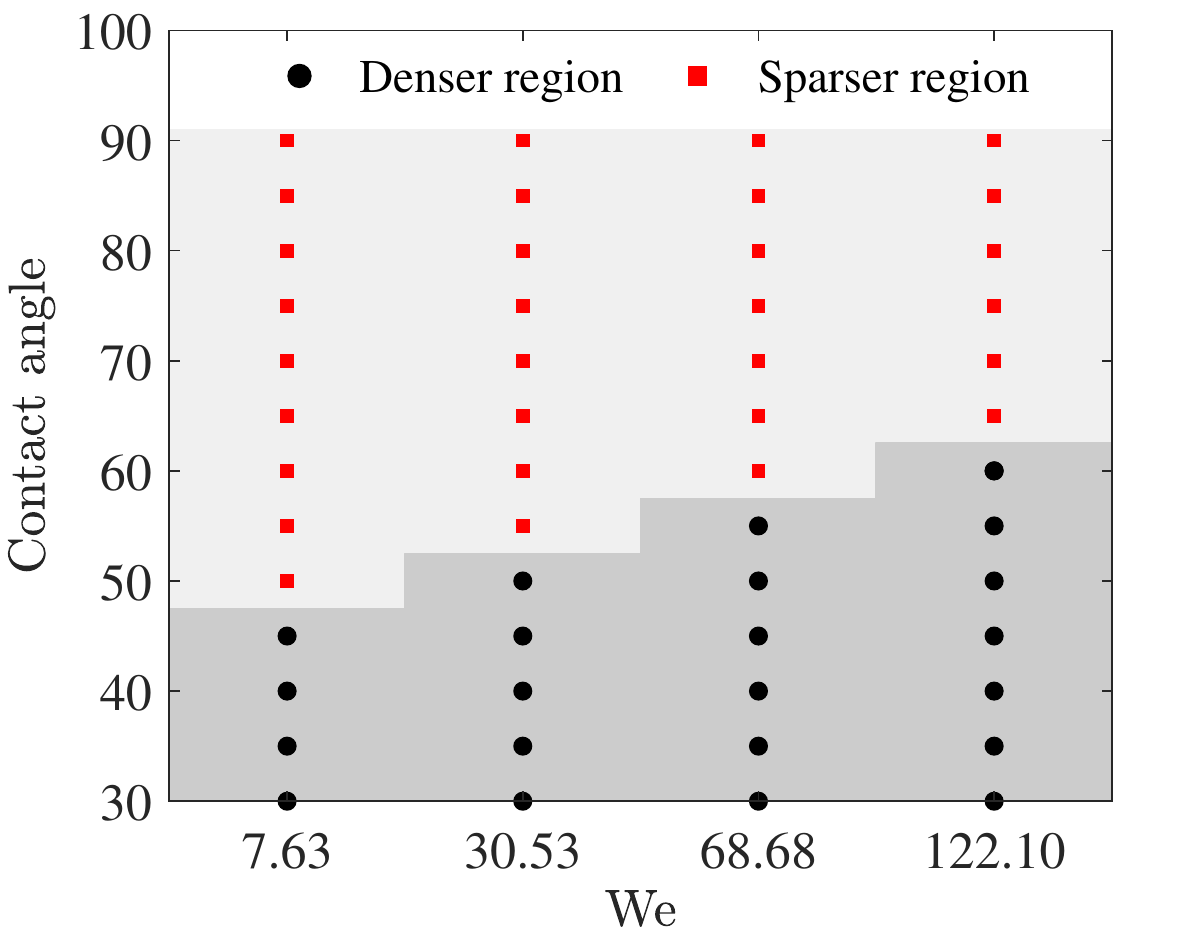}
		\caption{Phase diagram of the migration direction of a droplet as a function of contact angle and Weber number.}
		\label{fig17}
	\end{figure}

The above simulation results demonstrate the contact angle and Weber number play an important role in determining the migration direction of droplets at a given Janus-textured surface temperature. To elucidate the combined effect of Weber number and contact angle on migration direction, Fig. \ref{fig17} shows the phase diagram of the migration direction of droplets with different Weber numbers and contact angles. It can be seen that for large wettability, the droplets completely migrate to the denser region under the action of unbalanced Young's force in the asymmetric retracting phase II so that the droplet is completely independent of the vapor pressure gradient, resulting in the droplets eventually remain in the denser region. The migration direction of the droplet gradually changes from the sparser region to the denser region with the Weber number increases for the case of $CA= 50^{\circ} $, $CA= 55^{\circ} $ and $CA= 60^{\circ} $. However, the droplets move spontaneously toward the sparser region owning to the vapor pressure gradient for large contact angles.

\section{Conclusion}

In this work, the droplet impact dynamics on the Janus-textured heated surface are numerically investigated by the pseudopotential LB model. The numerical model is validated through the Laplace's law test, the contact angle test, and the droplet evaporation on the heated surface. The influences of the surface wettability, Jakob number and Weber number are all considered for the dynamic behavior of the impinging droplet on the Janus-textured heated surface. Based on the simulation result, the main conclusions are as follows.

\begin{itemize}
  \item [1)] 
  In the simulation, there are three different types of boiling states, i.e., the contact boiling state, the transition boiling state and the film boiling state, and the process of bubble nucleation, growth, and coalescence are captured.       
  \item [2)]
  The unbalanced Young's force exists in the lateral direction due to the wettability difference caused by the asymmetric structure. With this unbalanced force, the impinging process can be divided into three phases, i.e., the asymmetric spreading phase I, asymmetric retracting phase II and wetting equilibrium phase III.
  \item [3)]
  The Janus-textured surface with low wettability has superior nucleation behavior resulting in their typically easy transition from efficient transition boiling state to inefficient film boiling state. Meanwhile, the boiling states of the droplet change from the contact boiling state to the transition boiling state and then to the film boiling state as the Jakob number increases.
  \item [4)]
  The directional transport of the droplet is mainly controlled by the unbalanced Young's force, vapor pressure gradient and thermophoretic force on both sides of the Janus-textured surface. Under the action of these three forces, the droplet always moves toward the denser region in the transition boiling state while the droplet in the film boiling state transport to the sparser region.
  \item [5)]
  The contact time of droplets in the contact boiling state and transition boiling state is shorter than that in the film boiling regime. In addition, The droplets in the transition boiling state have a large evaporation rate due to the fact that the droplet interface is severely deformed by the evaporation and nucleation process, which greatly enhances the nucleate boiling heat transfer performance. However, the heat transport performance of the droplets in film boiling is poor due to the vapor layer.
\end{itemize}

\section*{Conflict of interest}
	We declare that we have no financial and personal relationships with other people or organizations that can inappropriately influence our work, there is no professional or other personal interest of any nature or kind in any product, service and/or company that could be construed as influencing the position presented in, or the review of, the manuscript entitled.

\section*{Acknowledgements}
	This work is financially supported by the National Natural Science Foundation of China (Grant No. 12002320), and the Fundamental Research Funds for the Central Universities (Grant Nos. CUG180618 and CUGGC05).


\end{document}